\documentclass[]{elsart}
\usepackage{graphicx}
\usepackage{amsmath, amsfonts, amssymb, bm}
\usepackage[]{psfrag}
\psfragscanoff

\def\im{i}
\def\figref{Fig. }
\def\eqnref{Eq. }
\def\eqnsref{Eq. }
\def\wnorm{\: [\gamma_1]}

\begin{document}

\runauthor{Evers, Bullock and Keitel}

\begin{frontmatter}
\title{Dark state suppression and narrow fluorescent feature in a laser-driven $\Lambda$ atom}
\author{J\"org Evers\thanksref{evers}},
\author{David Bullock} and
\author{Christoph H. Keitel\thanksref{keitel}}
\address{Theoretische Quantendynamik, Fakult\"at f\"ur Physik, Universit\"at Freiburg, Hermann-Herder-Stra{\ss}e 3, D-79104 Freiburg, Germany}
\thanks[evers]{E-mail: evers@physik.uni-freiburg.de}
\thanks[keitel]{E-mail: keitel@physik.uni-freiburg.de}
\begin{abstract}
We discuss quantum interference effects in a three-level atom in $\Lambda$-configuration, where both transitions from the upper
state to the lower states are driven by a single monochromatic laser field. Although the system has two lower states,
quantum interference is possible because there are interfering pathways to {\it each} of the two lower states.
The additional interference terms allow for interesting effects such as the suppression of a dark state which is present
without the interference. Finally we examine a narrow spectral feature in the resonance fluorescence of the atom with quantum
interference.
\end{abstract}

\end{frontmatter}

\section{Introduction}
It is well known that an atomic three-level system in V-configuration driven by laser fields gives rise
to many interesting quantum optical effects \cite{vsystem,scully,interference,qbeats,darkstates,dark2,plenio,darkperiods,v-resonance,zhou97,dress-v}.
If the two atomic upper states are closely-spaced such that the two transitions to the ground state interact with the same vacuum modes,
quantum interference may occur. In this case, there are more than one pathways to the lower state which can not be distinguished in
principle \cite{interference}. One consequence of this is
a possible modification of the resonance fluorescence spectrum of the atom allowing e.g. for quantum beat oscillations \cite{qbeats},
dark states \cite{darkstates,dark2}, macroscopic dark periods \cite{plenio,darkperiods} or very narrow resonances \cite{vsystem,dark2,v-resonance,zhou97}.

The atomic three level system may also be in $\Lambda$-configuration \cite{lambda}. While this system has been studied as intensively as
the V-system in most configurations, interference effects in the $\Lambda$-system have received less attention. One reason for
this might be that in this system, interference is not so obvious due to the fact that there are two lower states as possible
final states. Also it is a strong restriction that both in the V- and in the $\Lambda$-system the transitions have to interact with
the same vacuum modes for interference effects to occur. In practice, this condition is difficult to meet. For the V-system,
a dressed state realization was found some years ago using an atomic two-level system where one of the two states
is coherently coupled to an auxiliary level \cite{dress-v}. This makes the system much easier to access in practice. A corresponding
realization for the $\Lambda$-system was only recently found \cite{dressed}.
In \cite{javanainen}, the response of a $\Lambda$-system with interference to
a strong driving field was studied. It was shown that a dark state which is present in the system without interference \cite{darklambda} may vanish
with interference in the limit of high driving
intensity. \cite{dressed} demonstrates that the $\Lambda$-system with maximum quantum interference can be realized using
a dressed state picture of a coherently driven V-system without interference. This equivalence may be used to physically
interpret the features of the system. \cite{pumpprobe} examined the pump-probe response of the $\Lambda$-system, where interference
allows for quantitative changes.

In this paper we give a physical interpretation for the interference in the $\Lambda$-system. We show that the dark state may
also be suppressed if the driving field intensity is low. Finally we discuss a narrow spectral feature found in the resonance
fluorescence spectrum of the atom with interference utilizing the equivalence of the system to the corresponding
dressed state representation.

\section{System of interest}
\begin{figure}[t]
\centering
\psfrag{O}[][l]{$\bm \Omega$}
\psfrag{g1}{$\bm \gamma _1$}
\psfrag{g2}{$\bm \gamma _2$}
\psfrag{1}{$\bm |a\rangle$}
\psfrag{2}{$\bm |b\rangle$}
\psfrag{3}{$\bm |c\rangle$}
\psfrag{d}{$\bm \delta$}
\psfrag{D}{$\bm \Delta$}
\includegraphics[height=5cm]{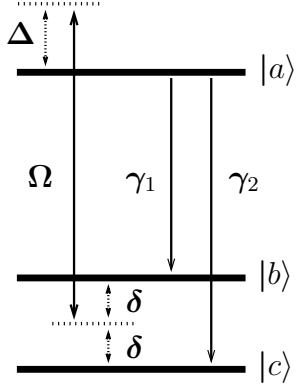}
\caption{\label{pic-system} The system of interest. The two lower levels $|b\rangle$ and $|c\rangle$ are closely spaced.}
\end{figure}

The system of interest is shown in \figref \ref{pic-system}. $\hbar \omega_i$ ($i=a,b,c$) are the respective energies of the three levels,
$\gamma_i$ ($i=1,2$) are the decay constants. Our aim is to study interference effects in the given system. Therefore we assume
the two lower levels $|b\rangle$ and $|c\rangle$ to be closely-spaced with
$2 \delta < \gamma_1, \gamma_2$ where $2\delta = \omega_b - \omega_c$. Both transitions $|a\rangle \leftrightarrow |b\rangle$ and
$|a\rangle \leftrightarrow |c\rangle$ are driven by a laser field of frequency $\omega_L$ with the respective Rabi frequencies $\Omega_1$
and $\Omega_2$. $\Delta = \omega_L - \omega_a + \omega_b + \delta$ is the common detuning of the driving laser field.

The system Hamiltonian is
\begin{eqnarray*}
H &=& H_0+H_1 \qquad \textrm{where}\\
H_0 &=& \hbar (\Delta + \omega _a) |a\rangle \langle a| + \hbar (- \delta + \omega _b) |b\rangle \langle b|   + \hbar ( \delta + \omega _c) |c\rangle \langle c|\\
H_1 &=& -\hbar \Delta |a\rangle \langle a| + \hbar \delta |b\rangle \langle b| - \hbar \delta |c\rangle \langle c|   \\
 && - \hbar \Omega_1 (e^{-i\omega_L t} |a\rangle \langle b|  + h.c. )   - \hbar \Omega_2 (e^{-i\omega_L t} |a\rangle \langle c|  + h.c. ).
\end{eqnarray*}

The interaction picture Hamiltonian with respect to $H_0$ is
\begin{equation*}
V = -\hbar \Delta |a\rangle \langle a| + \hbar \delta |b\rangle \langle b| - \hbar \delta |c\rangle \langle c|
 - \hbar \Omega_1 (|a\rangle \langle b|  + |b\rangle \langle a| )   - \hbar \Omega_2 (|a\rangle \langle c| + |c \rangle \langle a|).
\end{equation*}

This yields as equations of motion for the elements of the density matrix describing the system under the assumption that
the two lower levels are closely-spaced \cite{dressed,javanainen}
\begin{subequations}
\label{eoms}
\begin{eqnarray}
\dot{\rho} _{aa} &=& - (\gamma_1 + \gamma_2)\rho_{aa} - i\Omega_1 (\rho_ {ab} - \rho_{ba}) - i\Omega_2 (\rho_ {ac} - \rho_{ca})\\
\dot{\rho} _{bb} &=& \gamma_1 \rho_{aa} + i\Omega_1 (\rho_ {ab} - \rho_{ba}) \\
\dot{\rho} _{ab} &=& \left (i (\delta + \Delta) - \frac{1}{2}(\gamma_1 + \gamma_2)\right ) \rho_{ab} - i\Omega_1 (\rho_{aa} - \rho_{bb}) + i\Omega_2\rho_{cb} \label{eom}\\
\dot{\rho} _{ac} &=& \left (i (- \delta + \Delta) - \frac{1}{2}(\gamma_1 + \gamma_2)\right ) \rho_{ac} - i\Omega_2 (\rho_{aa} - \rho_{cc}) + i\Omega_1\rho_{bc} \\
\dot{\rho} _{bc} &=& p\sqrt{\gamma_1 \gamma_2}\rho_{aa} - 2i\delta \rho_{bc} +i\Omega_1 \rho_{ac} -i\Omega_2 \rho_{ba}
\end{eqnarray}
\end{subequations}
with $\rho_{ij} = (\rho_{ji})^*$ for $i,j \in \{a, b, c\}$ and $\rho_{aa}+\rho_{bb}+\rho_{cc}=1$.

$p$ is a measure for the quantum interference exhibited by the system. It depends on the value of $\delta$ and on the relative
direction of the two transition dipole elements. In the case of maximum interference with parallel dipole moments $p$ is equal to unity,
while $p=0$ represents the case of no quantum interference.

The equations of motion can be written as
\begin{eqnarray*}
\frac{d}{dt}\vec{\rho}  &=& \bm B \: \vec{\rho} + \vec{I} \qquad \textrm{with} \qquad \vec{\rho} = \langle \vec{\sigma} \rangle \qquad \textrm{and} \\
 \vec{\sigma} &=& ( |a\rangle\langle a|, |c\rangle\langle c|, |a\rangle\langle b|, |b\rangle\langle a|, |a\rangle\langle c|,
 |c\rangle\langle a|,  |b\rangle\langle c|, |c\rangle\langle b|)^T
\end{eqnarray*}
where $\bm B$ is a time independent time evolution matrix and $\vec{I}$ is a constant vector, both given by \eqnsref (\ref{eoms}).
In the following, we denote the $i$th element of the vector $\vec{\sigma}$ as $\sigma _i$.

To calculate the resonance fluorescence spectrum we need to calculate the two time correlation function
$\langle D^{(+)} (t) D(t')\rangle$ \cite{spectrum} given as
\begin{equation*}
D^{(+)}(t) = d_{ab} \sigma_3(t) + d_{ac} \sigma_5(t) \qquad \textrm{and} \qquad D^{(+)}(t) = (D^{(-)}(t))^{\dagger}.
\end{equation*}
$d_{ab}$ and $d_{ac}$ are the respective transition dipole moments with $|d_{ab}|^2 \propto \gamma_1$, $|d_{ac}|^2 \propto \gamma_2$ and
$d_{ab}d_{ac} \propto p\sqrt{\gamma_1 \gamma_2}$.
Splitting up $\vec{\sigma}$ into its average and the deviation from the average, $\sigma _i = \langle \sigma_i \rangle + \delta \sigma_i$
($i\in\{1,\dots,8\}$),
we obtain four terms corresponding to the average motion and four terms corresponding to the fluctuations. The average
motion yields the coherent part of the fluorescence spectrum
\begin{equation}
S_{coh}(\omega) = I_{coh}^{abs} \delta(\omega - \omega_L)  \label{coh-spec}
\end{equation}
where
\begin{equation}
I_{coh}^{abs} = \pi(\gamma_1 \rho_{ab}^{ss}\rho_{ba}^{ss} + \gamma_2 \rho_{ca}^{ss}\rho_{ac}^{ss} +  p \sqrt{\gamma_1 \gamma_2}\rho_{ab}^{ss}\rho_{ca}^{ss} + p \sqrt{\gamma_1 \gamma_2}\rho_{ac}^{ss}\rho_{ba}^{ss}) \label{int-coh}
\end{equation}
is the absolute intensity of the coherent peak. The superscripts ''ss'' denote the steady state values of the system.

Using the quantum regression theorem, the normalized incoherent fluorescence spectrum can be
calculated from the four fluctuation terms of the splitted $\vec{\sigma}$ as
\begin{equation}
S_{inc}(\omega) = \frac{1}{\pi \rho^{ss}_{aa}} \text{Re}\: \{ {\bm K}_3 (\gamma_1 \vec{R}_{ab} + p \sqrt{\gamma_1 \gamma_2}\vec{R}_{ac})
 +{\bm K}_5(\gamma_2 \vec{R}_{ac} + p \sqrt{\gamma_1 \gamma_2}\vec{R}_{ab}) \}
 \label{inel-spec}
\end{equation}
with
\begin{eqnarray*}
{\bm K} &=& (\im \omega - {\bm B})^{-1}, \: \vec{R}_{ab} = \langle \delta \vec{\sigma}(0) \delta \sigma_{4}(0)\rangle^{ss}
\: \textrm{and} \: \vec{R}_{ac} = \langle \delta \vec{\sigma}(0) \delta \sigma_{6}(0)\rangle^{ss}.
\end{eqnarray*}

The total fluorescence intensity is given by the expression
\begin{equation*}
I_{tot} = \pi (\gamma_1 + \gamma_2) \rho^{ss}_{aa}
\end{equation*}
such that in the following attention will be placed on both the various spectral components and the upper level population $\rho^{ss}_{aa}$.

\section{\label{sec-spec} Investigation of the spectra}
In this section we calculate the resonance fluorescence spectra emitted by the given system. In the first part we use the results of
\eqnref (\ref{inel-spec}) for a numerical analysis of the incoherent contribution to the spectrum. In the second part, we discuss some
of the system properties using analytical arguments. The elastic component of the spectrum given by \eqnref (\ref{coh-spec}) is
omitted in all figures. The numerical values in this section are given in units of the decay constant $\gamma_1$.

\begin{figure}[t]
\centering
\psfrag{xlbl}[tc][c]{$\omega-\omega_L \wnorm$}
\psfrag{ylbl}[ct][c][1][180]{rel. intensity}
\psfrag{pa}{(a)}
\psfrag{pb}{(b)}
\includegraphics[height=3.5cm]{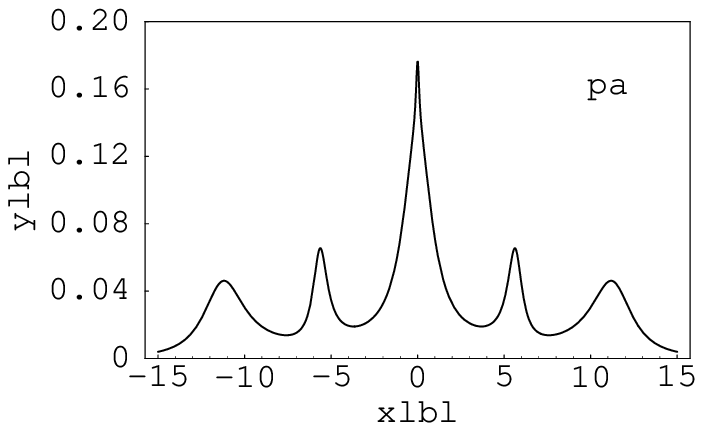}
\hspace{0.4cm}
\includegraphics[height=3.5cm]{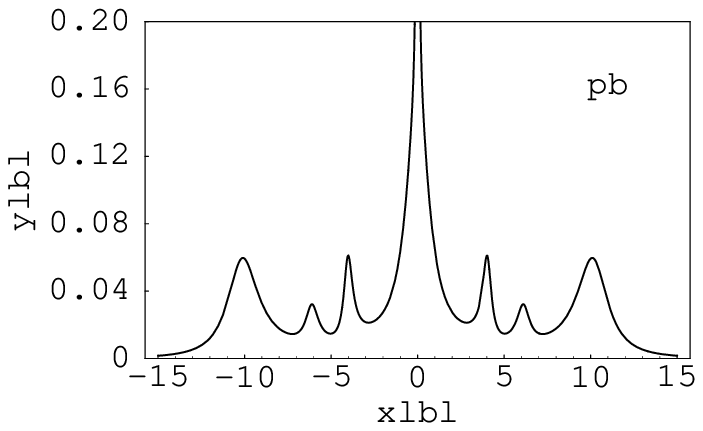}
\caption{\label{pic-sample} Normalized inelastic spectra. (a) $\gamma_1 = \gamma_2 = 1$, $\Omega_1 = \Omega_2 = 4$,
$\Delta = 0$, $p=0.8$ and $\delta = 0.5$. (b) $\gamma_1 = 1$, $\gamma_2 = 1$, $\Omega_1 = 3$, $\Omega_2 = 4$,
$\Delta = 2$, $p=1$ and $\delta = 0.1$.}
\end{figure}

\begin{figure}[t]
\centering
\psfrag{xlbl}[tc][c]{$\omega-\omega_L \wnorm$}
\psfrag{ylbl}[ct][c][1][180]{rel. intensity}
\psfrag{pa}{(a)}
\psfrag{pb}{(b)}
\includegraphics[height=3.5cm]{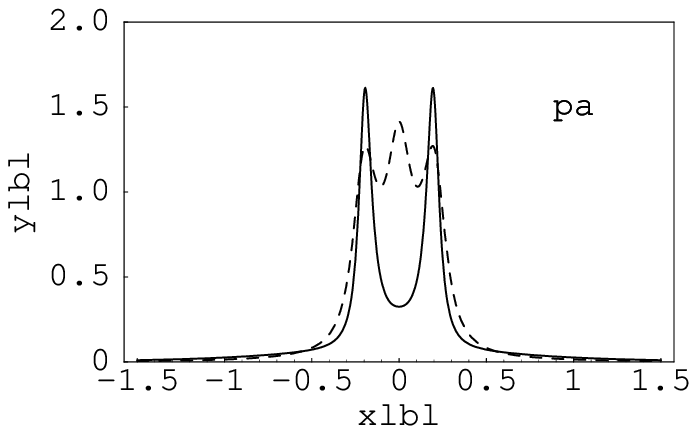}
\hspace{0.4cm}
\includegraphics[height=3.5cm]{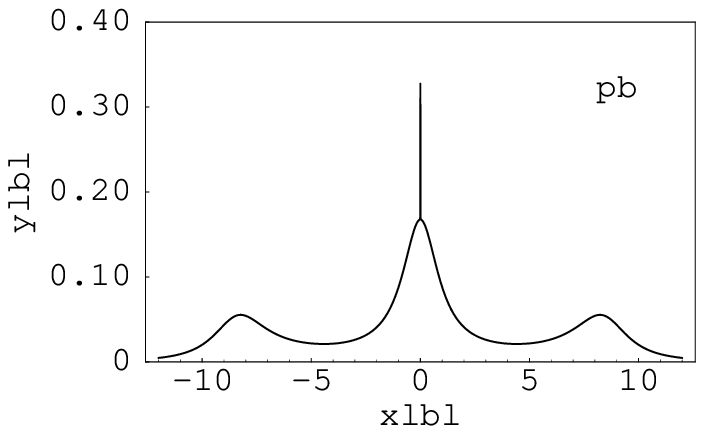}
\caption{\label{pic-peak} Normalized inelastic spectra. (a) $\gamma_1 = \gamma_2 = 1$, $\Omega_1 = \Omega_2 = 0.2$,
$\Delta = 0$ and $\delta = 0.1$. Solid line: $p=1$, dashed line: $p=0$.
(b) exhibits the additional narrow peak centered at the laser frequency.
The parameters are chosen as $\gamma_1 = \gamma_2 = 1$, $\Omega_1 = \Omega_2 = 3$,
$\Delta = 0$, $p=1$ and $\delta = 0.1$.}
\end{figure}

A sample spectrum is shown in  \figref \ref{pic-sample}(a).
The spectrum consists of a middle structure centered
at the driving laser field frequency and two pairs of sidebands. In general, the fluorescence spectrum consists of 9 structures:
The coherent peak, two incoherent peaks at the driving laser field frequency and three pairs of sidebands. The various incoherent
contributions can be distinguished in \figref \ref{pic-sample}(b). However the absolute intensity for this choice of parameters is very low.
Depending on the chosen parameters, some of these structures may be suppressed. For example in \figref \ref{pic-sample}(a) there is only
one pair of sidebands for $p=1$ instead of $p=0.8$. Another case is shown in \figref \ref{pic-peak}(a)
. The solid curve is calculated for maximum interference ($p=1$), while the dashed line is plotted
without interference term ($p=0$). In this parameter range, the middle structure is only present without interference.

The parameters may also be chosen such that the system exhibits an additional narrow peak centered at the frequency of the driving laser field.
This is shown in \figref \ref{pic-peak}(b)
. In this figure, the relative amplitude of the fluorescence spectrum at the driving laser frequency is $A \approx 0.365$, and the
width of the additional narrow peak is $\Gamma \approx 0.0011$. The narrow peak vanishes with increasing Rabi
frequency if the other parameters are kept fixed. However by adjusting the detuning $\Delta$ appropriately with the Rabi frequency, it is possible to have a narrow peak even
for high Rabi frequencies \cite{narrowfeature}. An approximate value for the optimal detuning $\Delta_{max}$ may be obtained by maximizing the expression in
\eqnref (\ref{int-coh}) for the absolute intensity of the coherent contribution to the fluorescence spectrum with respect to $\Delta$.
For $\gamma_1 = \gamma_2 = \gamma$, $\Omega_1 = \Omega_2 = \Omega$ and $p=1$ this yields $\Delta_{max} = \pm \sqrt{2\Omega^2 + \delta^2 -
\gamma^2}$. With this detuning and the other parameters as in \figref \ref{pic-peak}(b), the amplitude of the fluorescence spectrum at the
driving laser frequency becomes $A \approx 17.951$, and the width of the narrow peak is $\Gamma \approx 0.0015$. Thus in this case, the
relative intensity of the additional peak is more than 220 times larger than with zero detuning.

Of course it is of great interest to find consequences of the quantum interference accessible to experiments. There are parameter 
ranges where the system of interest exhibits a dark state \cite{darklambda}. In such a dark state,
the steady state population of the upper atomic level and thus the total fluorescence intensity vanishes. The dark state may be 
destroyed by a finite laser bandwidth \cite{bandwidth} or by applying thermal or squeezed fields \cite{thermal}. For these schemes, quantum interference is not needed. However as pointed 
out in \cite{dressed,javanainen}, the dark state
may also be suppressed by quantum interference such that the system is in a dark state without interference  $(p=0)$, but not 
with interference $(p=1)$. For this, the Rabi frequency does not need to be very high if the two lower levels are close to each 
other. For example, for $\gamma_1 = \gamma_2 = \gamma$ and $\Omega_1 = \Omega_2 = \Omega$ we have with maximum interference ($p=1$)
\begin{align}
\rho^{ss}_{aa} &= \frac{\Omega ^2}{\Delta ^2 + \delta ^2 + \gamma ^2 + 2\Omega ^2} \xrightarrow{\delta \rightarrow 0} \frac{\Omega ^2}{\Delta ^2 + \gamma ^2 + 2\Omega ^2}.\\
\intertext{Without interference ($p=0$), the corresponding population is}
\rho^{ss}_{aa} &= \frac{\delta ^2 \Omega ^2}{\delta ^2 \Delta ^2 + \delta ^4 + \Omega ^4 + \delta ^2 (\gamma ^2 + \Omega ^2)} \xrightarrow{\delta \rightarrow 0} 0.
\end{align}
Thus for $\delta \rightarrow 0$, the system is in a dark state without the interference, but not with quantum interference. Without interference,
the dark state results from coherent population trapping as for $\delta \rightarrow 0$ both transitions are driven resonantly.
The suppression of the dark state is crucial for the narrow peak, as it is most pronounced for low values of $\delta$ such that the total
intensity becomes low without the interference. 

Using the fact that the system studied here with full quantum interference ($p=1$) is equivalent to an appropriately chosen three-level
system in V-configuration without quantum interference \cite{dressed}, the results for the narrow peak for such a V-system in
\cite{plenio} can be rewritten to analytically describe the narrow peak in the $\Lambda$-system with interference. With
$\gamma_1 = \gamma_2 = \gamma$ and
$\Omega_1 = \Omega_2 = \Omega$ we obtain as approximate expressions for the full width at half height $\Gamma$ and the relative height $A$
for the peak:
\begin{eqnarray}
\Gamma &=& 2\gamma \: \frac{\delta^2}{\Omega^2} \: \frac{\Delta^2+\gamma^2+2\Omega^2}{\Delta^2+\gamma^2+4\Omega^2}\\
A &=& \frac{1}{\pi}\: \frac{\Omega^2}{\delta^2}\: \frac{(\Delta^2+\gamma^2)^2}{4\gamma (\Delta^2 + \gamma^2 + 2\Omega^2)^2}
\end{eqnarray}
These analytical expressions are in good agreement with the numerical results. As in the corresponding V-system with interference
\cite{zhou97}, the width is proportional to $\gamma \cdot (\delta / \Omega)^2$. But the height $A$
is inversely proportional to this factor in both our and the corresponding V-system. Thus the relative intensity
\begin{equation}
\Gamma \cdot A = \frac{(\Delta^2+\gamma^2)^2}{2\pi (\Delta^2 + \gamma^2 + 2\Omega^2)(\Delta^2+\gamma^2+4\Omega^2)}
\end{equation}
is independent of the energy splitting $\delta$ of the two lower levels as long as it is small enough to account for
quantum interference. However in the corresponding V-system with interference, the total fluorescence intensity completely vanishes
for $\Delta=0$ and $p=1$ in our notation \cite{zhou97,exp-vanish}. The atom only emits fluorescence light if $p$ is slightly less
than unity. This effect is not present in our system.

The equivalence of the system discussed here with a corresponding system without interference may also
be used to explain why an appropriately changed detuning may increase the peak intensity as in \cite{narrowfeature}.

\section{Physical interpretation}
\begin{figure}[t]
\centering
\psfrag{1}{$\bm |a\rangle$}
\psfrag{2}{$\bm |b\rangle$}
\psfrag{3}{$\bm |c\rangle$}
\includegraphics[height=3cm]{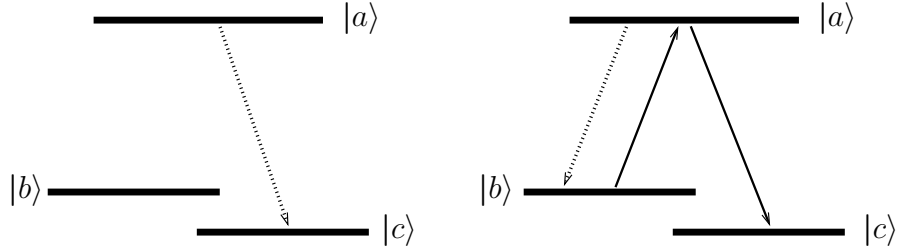}
\caption{\label{pic-paths} Two interfering paths from $|a\rangle$ to $|c\rangle$. The dashed line denotes spontaneous decay, the solid
lines denote laser-induced transitions.}
\end{figure}

As compared to a three-level system in $\Lambda$-configuration with well-separated lower levels, the system presented here has a distinctively
different behavior for several parameter ranges due to interference effects.
Quantum interference occurs in systems with transitions for which there are at least two different ways to one final state which cannot be
distinguished in principle. A typical example for this is the two-slit experiment \cite{qbeats}. In an atomic three-level system in V-configuration with
closely-spaced upper levels, quantum interference is possible because the energy arising from the decay of one of the upper levels
may excite the atom into the other upper level due to the coupling through the same vacuum modes. This introduces an extra
coherence between the two upper states. A driving laser field may couple to this coherence and thus account for interference effects also
in the resonance fluorescence spectrum. In the $\Lambda$-system considered here, the possibility for quantum interference is not so
obvious. However there are interfering pathways to {\it each} of the final states $|b\rangle$ and $|c\rangle$. Also in this system,
there is an extra coherence term between the two lower levels due to the spontaneous decay from the upper level. The driving laser field
couples to this coherence and allows for additional paths for the transitions from the upper to the lower levels. Two interfering
paths for transition $|a\rangle \rightarrow |c\rangle$ are shown in \figref \ref{pic-paths}. The laser-induced transition
from $|b\rangle$ to $|c\rangle$ is possible because the energy difference $2\delta$ between the two lower states is within the energy
uncertainty of the two transitions. The interfering pathways with final state $|b\rangle$ are equivalent to the ones shown in
\figref \ref{pic-paths}.

Due to the interference, the dark state usually found in a $\Lambda$-system without interference (\cite{plenio}) may be suppressed.
This can be used to find parameter ranges where the additional narrow peak is especially narrow and intense.
The various other spectral features found in section \ref{sec-spec} may be explained using a dressed state picture such as described
in \cite{zhou97} for a three-level system in V-configuration.

\section{Discussion and Conclusion}
We have calculated and discussed the resonance fluorescence spectrum of an atomic three-level system in $\Lambda$-configuration
where the two lower states are closely-spaced. Both transitions are driven by a
laser field. The two transitions couple to the same vacuum modes such that quantum interference may occur.
Due to the additional interference terms, the dark state usually found in a $\Lambda$-system can be suppressed. This allows
to choose the system parameters favorably for an additional narrow peak centered at the laser frequency. Although quantum
interference is not so obvious in the given system because of the fact that there are two final states, the other spectral features are
found to be analogous to the spectra of a laser-driven three-level system in V-configuration with interference between the two upper states.

\ack
Funding by Deutsche Forschungsgemeinschaft (Nachwuchsgruppe within SFB 276) is gratefully acknowledged.


\begin{thebibliography}{999}
\bibitem{vsystem} L.M. Narducci, M.O. Scully, G.L. Oppo, P. Ru, and J.R. Tredicce, Phys. Rev. A 42 (1990) 1630;
D.J. Gauthier, Y. Zhu and T.W. Mossberg, Phys. Rev. Lett. 66 (1991) 2460.
\bibitem{scully} M. O. Scully and S. Y. Zhu, Science 281 (1998) 1973.
\bibitem{interference} Z. Ficek, {\it Quantum interference in atomic and molecular systems}, Advances in chemical physics 119, part 1,
Wiley, New York (2001) and references therein.
\bibitem{qbeats} P. Meystre and M. Sargent III, {\it Elements of quantum optics}, Springer-Verlag, Berlin (1999); M. O. Scully, Phys. Rev. Lett. 55 (1985) 2802.
\bibitem{darkstates} D. A. Cardimona, M. G. Raymer and C. R. Stroud Jr., J. Phys. B 15 (1982) 65.
\bibitem{dark2} S. Y. Zhu, R. C. F. Chan and C. P. Lee, Phys. Rev. A 52 (1995) 710; S. Y. Zhu, L. M. Narducci and M. O. Scully, Phys. Rev. A 52 (1995) 4791.
\bibitem{plenio} G. C. Hegerfeldt and M. B. Plenio, Phys. Rev. A 52 (1995) 3333.
\bibitem{darkperiods} B. M. Garraway, M. S. Kim and P. L. Knight, Opt. Commun. 117 (1995) 550.
\bibitem{v-resonance} P. Zhou and S. Swain, Phys. Rev. Lett. 77 (1996) 3995; narrow controllable structures were
shown also in related four-level systems by E. Paspalakis and  P. L. Knight, Phys. Rev. Lett. 81 (1998) 293;
C. H. Keitel, Phys. Rev. Lett. 83 (1999) 1307.
\bibitem{zhou97} P. Zhou and S. Swain, Phys. Rev. A 56 (1997) 3011.
\bibitem{dress-v} M. Fleischhauer, C. H. Keitel, L. M. Narducci, M. O. Scully, S. Y. Zhu and M. S. Zubairy, Opt. Commun. 94 (1992) 599.
\bibitem{lambda} A. S. Manka, H. M. Doss, L. M. Narducci, P. Ru and G. L. Oppo, Phys. Rev. A 43 (1991) 3748; G. S. Agarwal, {\it ibid.} 54 (1996) R3734.
\bibitem{dressed} X.-M. Hu and J.-S. Peng, J. Phys. B 33 (2000) 921.
\bibitem{javanainen} J. Javanainen, Europhys. Lett. 17 (1992) 407.
\bibitem{darklambda} G. Alzetta, A. Gozzini, L. Moi and G. Orriols, Nuovo Cimento 36B (1976) 5;
E. Arimondo and G. Orriols, Lettre al  Nuovo Cimento 17 (1976) 333; R. M. Whitely and C. R. Stroud, Phys. Rev. A 14 (1976) 1498.
\bibitem{pumpprobe} S. Menon and G. S. Agarwal, Phys. Rev. A 57 (1998) 4014.
\bibitem{spectrum} C. Cohen-Tannoudji, {\it Atoms in Strong Resonant Fields}, in: Frontiers in Laser Spectroscopy, Les Houches Summerschool XXVII (1975), North-Holland Publ. Co., Amsterdam  (1977).
\bibitem{bandwidth} B. J. Dalton and P. L. Knight, J. Phys. B 15 (1982) 3997.
\bibitem{thermal} M. R. Ferguson, Z. Ficek and B. J. Dalton, Phys. Rev. A 54 (1996) 2379.
\bibitem{exp-vanish} H. R. Xia, C. Y. Ye and S. Y. Zhu, Phys. Rev. Lett. 77 (1996) 1032.
\bibitem{narrowfeature} J. Evers and C. H. Keitel, Phys. Rev. A 65 (2002) 033813.
\end{thebibliography}
\end{document}